# SECURITY OF PLUG-AND-PLAY QKD ARRANGEMENTS WITH FINITE RESOURCES

PEDRO J. SALAS[a]

*Departamento Tecnologías Especiales Aplicadas a la Telecomunicación, Universidad Politécnica de Madrid, E.T.S.I. Telecomunicación, Ciudad Universitaria s/n, 28040 Madrid, Spain*

The security of a passive plug-and-play QKD arrangement in the case of finite (resources) key lengths is analysed. It is assumed that the eavesdropper has full access to the channel so an unknown and untrusted source is assumed. To take into account the security of the BB84 protocol under collective attacks within the framework of quantum adversaries, a full treatment provides the well-known equations for the secure key rate. A numerical simulation keeping a minimum number of initial parameters constant as the total error sought and the number of pulses is carried out. The remaining parameters are optimized to produce the maximum secure key rate. Two main strategies are addressed: with and without two-decoy-states including the optimization of signal to decoy relationship.

## 1 Introduction

The need to communicate messages secretly is in the roots of humanity. To provide secure communications without the flaw of being potentially insecure as technology improves, the concept of unconditional security has been introduced. This security is based on mathematical properties instead of the present day adversary's abilities. As is well known, the Vernam cipher fulfils the unconditional security while the secret key is not reused. As a consequence, the cryptographic protocol needs a secure source of fresh keys to be distributed between the interlocutors (usually called Alice and Bob). The main goal for the QKD [1] protocols and set-ups is to provide these secret keys to the users. In this framework, decoy state method [2] brought about a significant improvement in the QKD performance. The main idea of decoy states is that Alice changes, at random, the characteristics of some extra pulses (decoy states) sent to Bob, revealing this information only at the end of the transmission. Therefore, the eavesdropper cannot adjust her attack to each pulse shared. This extra knowledge allows the interlocutors (in the post-processing step) to improve the estimations of the parameters involved in the key distribution task. This represents a way of mitigating the photon-number-splitting (PNS) attack on QKD protocols using weak laser pulses. If decoy-states are used with the BB84 protocol, the secure key rate is proportional to the overall transmittance, even for attenuated light, instead of the usual square dependence when no decoy states are used. As the channel transmission is usually quite low, this fact increases the key rate when decoy states are used. When an infinite number of decoy states are used, the interlocutors can accurately estimate the variables involved in the key rate. In practice, the two decoy states method (vacuum + weak decoy) is enough to provide good results.

Despite the security protocols are well understood, proofs about their security, although of crucial interest, has shown to be hard work. For QKD, in the asymptotic limit of very long key lengths, the unconditional security [3] has already been proven. Several strategies have been used, based on: the uncertainty principle [4], the entanglement distillation protocols [5] even with imperfects devices [6] and the information-theoretical techniques [7]. A complementarity scenario [8,9,10] has been used to provide the necessary and sufficient conditions for the key distillation.

---

[a] Permanent email address psalas@etsit.upm.es

In security proofs against collective attacks, Alice and Bob usually have to characterize the density matrix of their shared state. Some recent techniques provided a tighter bound with unconditional security using uncertainty relations [11]. Better key generation rates, valid for general coherent attacks that guarantee unconditional security, are presented in [12] for the BB84 protocol with finite key lengths.

Cryptographic primitives are often used as components of other protocols, so it is natural to require the security of these new schemes. The notion is captured by the universal security: a cryptographic primitive is universally secure if it is secure in any arbitrary context. Unfortunately some commonly-used security definitions do not fulfil this property.

A typical QKD scenario involves two phases: (i) the generation and distribution of quantum signals and (ii) a discussion between Alice (the emitter) and Bob (the receiver) through an authenticated classic channel to carry out some tasks such as sifting, error estimation, error correction and privacy amplification. If Eve (the eavesdropper) carries out an individual attack on the signals, the whole process can be characterized by means of a three-random variable probability $P(K_A, K_B, Z)$ representing the classic results for Alice and Bob's keys ($K_A$, $K_B$) and Eve's information ($Z$). Then, the interlocutors can see if a public discussion can transform the data into a secret key providing a positive key generation rate $R$ (here defined as the average number of final secure key bits from each initial pulse sent). In this case, the security condition could be inherited from a classic one, basically requiring that $K_A = K_B = K$ (except with a small probability $\varepsilon$), having the maximum Shannon entropy and sharing a minimum information with Eve, i.e. $\max_{\forall Z} I(K:Z) \leq \varepsilon$.

The key generation rate ($R$) fulfils two bounds: a) $\max_{\forall Z}(I(K_A : K_B) - I(K_A : Z), I(K_A : K_B) - I(K_B : Z)) \leq R$ (Csiszár-Körner [13]), and b) $R \leq I(K_A, K_B \downarrow Z))$ (Maurer [14]), ($I(*:*)$ being the mutual information and $I(K_A, K_B \downarrow Z)$ the intrinsic information). The lower bound (a) is carried out by taking the maximum with respect to all of Eve's possible attacks giving rise to the random variable $Z$. For this purpose, in a first step, Alice and Bob carry out an error correction by means of a one-way method, with either Alice providing error information (the first term in the equation) or vice versa (the second term in the equation). In the second step, Alice and Bob execute a privacy amplification protocol shortening the key length by a fraction $I(K_A:Z)$ or $I(K_B:Z)$, depending on the direction of the communication. The upper bound (b) is written by means of intrinsic information defined as:

$$I(K_A, K_B \downarrow Z) = \mathrm{Inf}_{\forall P(\overline{Z}|Z)} \left\{ I(K_A, K_B | \overline{Z}), P(K_A, K_B, \overline{Z}) = \sum_{z \in Z} P(K_A, K_B, Z) P(\overline{Z}|Z) \right\},$$

the infimum of $I(K_A, K_B | \overline{Z})$ taken over all possible conditional distributions $P(\overline{Z}|Z)$, $Z$ being the output of a channel characterized by the conditional probability $P(\overline{Z}|Z)$. The intrinsic information measures the information that Bob learns about Alice's information, after Eve has published her data. If Bob's information depends only on Eve's data, the intrinsic information vanishes and $R = 0$.

The aforementioned security definition, although reasonable, has significant weaknesses. The value of $\varepsilon$ cannot be zero because perfect security ($\varepsilon=0$) is impossible using a probabilistic QKD protocol or running for a finite time, i.e., providing finite key lengths. As a consequence, the definition does not guarantee the universal security of the key. The reason is that we cannot force Eve to carry out a measure to get classic information. Eve is free to keep the information stored in quantum states (quantum adversaries), generally until any later time convenient to her. This strategy is permitted in collective attacks. Although the value of $I(K:Z)$ were negligible small, it may be that the key were completely insecure [15]. A new definition of $\varepsilon$-security fulfilling the universal security condition has already been proposed and will be considered in section 2.3.

Just after introducing the concept of decoy states, some attempts were carried out to introduce the parameter fluctuation coming from using finite resources. In [16], the theoretical effect of the fluctuations appearing in a real-life QKD experiments is considered using several decoy protocols. Fluctuations are also taken into account in some experimental implementations as in [17]. The effect of source errors and statistical fluctuations in a 3-intensity decoy-state protocol is treated in [18]. The $\varepsilon$-security definition is assumed in some theoretical analyses. In [19], the effect of a finite key is studied although optimal error correction and no parameter fluctuation is assumed. Some improvement in the secret key rate is achieved by modifying the parameter estimation strategy in [20]. Simple arguments [21] allows to estimate that no secret-key could extracted if the number of pulses considered is smaller than $10^5$-$10^6$.

One of the most widely used set-ups to carry out QKD is the so-called "plug-and-play" arrangement. Its security is not so obvious because the device permits Eve to manipulate the pulses in any sophisticated way for her purposes. This fact gives rise to serious limitations to guarantee its security. In spite of some results have been achieved using the $\varepsilon$-security criterion in [22,23,24], they are not strictly applicable to these devices. The objective of this paper is to analyse the effect of the key finiteness, by means of the $\varepsilon$-security definition, when a plug-and-play set-up for QKD is used. Several situations, for infinite and finite key lengths will be dealt with and without decoy states and including weak + vacuum two-decoy states. In section 2, the main characteristics assumed for the set-up are compiled in addition to the theoretical description, both in the infinite as well as the finite key case including decoy states or not. Section 3 shows the results of the numerical simulation optimizing the main parameters.

## 2 Secure key rate for BB84 protocol

The original BB84 protocol involves a perfect 1-photon source, not completely available yet with the current technology. Instead of these sources, weakly coherent ones have been extensively used. In this context, the BB84 protocol has been demonstrated to be secure, by means of the tagged/untagged pulse distinction. The concept of tagged qubits (as having its information revealed to Eve, i.e. multi-photon pulses prone to photon-number-splitting attack) and untagged qubits (as secure to Eve's observation, i.e. 1-photon states), have been introduced in [6]. In the well-known GLLP method, post-processing takes advantage of this separation of pulses. In the first step, Alice and Bob sacrifice a fraction $h_2(E)$ of all qubits (or pulses) of the raw key ($E$ being the quantum bit error rate QBER) and $h_2$ the usual binary Shannon information function. The second post-processing step is to apply privacy amplification to the untagged qubits (pulses) because they are the secret fraction of the key. The secret key rate in the asymptotic limit of infinitely long keys is:

$$R^\infty = q \{-f(E) Q h_2(E) + Q_1 [1-h_2(E_1)]\},$$

$q$ being the sifting factor ($q=1/2$ for BB84) and $f(E)$ is the error correction inefficiency (usually $f(E) \sim 1$), $Q$ the overall gain and $Q_1$ and $E_1$ are the gain and error of untagged qubits. Unfortunately, the plug-and-play devices have their own peculiar security characteristics because the eavesdropper could tamper with the source and the channel. Therefore it is not allowed to suppose any specific light statistic, thus making the security analysis difficult. In spite of the above definition between tagged/untagged pulses being a good tool to discuss the security, their definition has to be slightly modified for these set-ups.

Together with a weak coherent light source, the squash model for Bob's detector is assumed [25,6]. In this model, the detection device is described by means of a two-step process: first, the photon signal is filtered and mapped (squashed) into a single photon state (qubit), i.e. a two-dimensional Hilbert space. If it succeeds, an ideal measurement is carried out; otherwise, the detector output shows a signal failure. The squashing model has been demonstrated for the BB84 protocol [26], even including the implementation of a passive basis choice [27], allowing security proofs [28] based on the assumption that Eve only sends one-photon states.

*2.1 The plug-and-play set-up*

One of the most widely used set-ups to carry out QKD is the so-called "plug-and-play" architecture. Bob (the final receiver) generates bright pulses that are sent to Alice (the final emitter) through the noisy channel connecting them. Alice encodes the information, attenuates the pulses, and sends them back to Bob. Eve has full control of the channel and, in the worst case, could replace any signal with a sophisticated one in order to achieve the best result (Trojan horse attack, [29]). To study the set-up security, the source is assumed to be completely unknown, as Eve can have a full control of it. This scenario is called QKD *with unknown and untrusted source* (UUS).

This situation has been analysed in detail in [30] for the case of infinite key lengths using an active method to sample the untrusted source and, in [31, 32], the authors propose a passive strategy (more appropriate to experimental implementation) to evaluate the secure key rate, providing an estimation of the effect of the finite key length, but without using a full treatment.

In this work, the main lines proposed in [31, 32] for a passive set-up, that are compiled in the following, will be assumed. Bob generates bright coherent pulses (with an average photon number per pulse $M_B$) that are sent to Alice through a noisy channel with transmittance $\eta$ (= $\eta_B$ $10^{-\beta L/10}$, $\eta_B$ = internal losses of Bob's device, $L$ = channel length and $\beta$ = channel loss coefficient in dB/km). Eve has full control of the source and the communication channel and, as a consequence, it is no longer correct to assume a Poissonian distribution for the photon number statistics of Bob's source, as is usually supposed in standard security proofs. Therefore, Alice should consider the source is completely unknown and untrusted (UUS). Alice has to introduce some strategies to improve the security of the whole process. The pulses received in Alice's set-up cross a filter to guarantee the single mode assumption, and a phase randomizer, transforming the pulse into a mixture of Fock states. The final secret key will be extracted only from the subset of pulses having a photon number $m$ in the interval $m \in [(1-\delta) M_A, (1+\delta) M_A]$, $\delta$ being a small positive number and $M_A$ is a parameter to be chosen by Alice and Bob, and can be identified as the average photon number per pulse of Alice's source. These pulses are defined as *untagged* and those out of this range are *tagged*. In order to obtain information on the photon number distribution of untagged pulses, Alice needs to monitor its energy. In the passive strategy, she uses a beam splitter ($q_A/1-q_A$) connected to an intensity detector (see the upper part of figure 1). Then, the pulses are sent to an encoder and an attenuator with transmittance $\lambda$. All of Alice's internal losses are modelled as a $\lambda/(1-\lambda)$ beam splitter, $0 \leq \lambda \leq 1$. Finally, the pulses are sent back to Bob. To calculate the final key generation rate $R$ by means of this passive strategy, Alice should estimate the fraction of untagged coding pulses from her measurements of the number of photon of each sampling pulse, using the intensity monitor. The probability bounds are also needed.

The estimation of the untagged pulse fraction is not as easy as one might think at first sight. In the splitting step, each pulse is separated into two: the U pulse sent to the encoder and the L pulse sent to the intensity monitor (see figure 1). One might argue that measuring the photon number of an L pulse could infer the photon number of its U partner. But this is not correct. For a pair (U, L) coming from the same pulse, the total number of photons is (constant but) unknown. Then, the number of photons of both pulses are correlated variables and do not fulfil the random sampling theorem. Fortunately, it can be demonstrated [32] that the number of untagged U pulses can be upper bounded to the number of untagged L pulses (measured by Alice's device) with a confidence parameter that depends on the total number of pulses.

Now we need to estimate the probability bounds to calculate $R$. This passive set-up is easily analysed by considering an equivalent (in the sense of providing the same probability bounds) active arrangement in which, instead of the beam splitter ($q_A/1-q_A$), another beam splitter ($q_A'/(1-q_A')) \equiv (1-q_A)/q_A$ (assuming the detector efficiency is the unit) and an optical active switch (50%) is used (see figure 1 and details in [31, 32]).

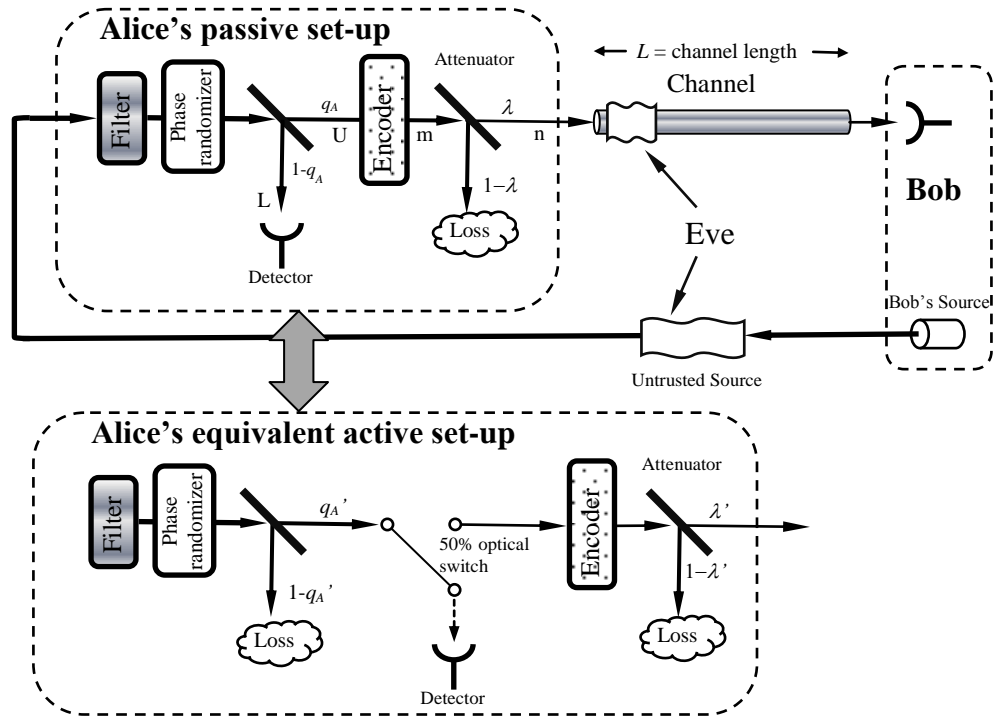

Fig.1. Alice's plug-and-play set-ups: passive arrangement and equivalent active arrangement

The starting point will be to characterize the unknown and untrusted source. Restricting the analysis to untagged pulses, makes the security analysis easier, as its narrow range of photon numbers allows us to obtain some upper and lower bounds for the photon distribution probabilities (see Eqs. 1a, b).

In spite of the source being completely unknown, as regards the untagged pulses, after Alice's attenuator, the conditional probability that $n$ photons are emitted by Alice given $m$ photon going in (see figure 1), follows the binomial distribution $P(n|m) = \binom{m}{n} \lambda^n (1-\lambda)^{m-n}$. Under the condition $(1+\delta) M_A \lambda < 1$ (reflecting an expected output photon number for untagged bits lower than 1), this distribution can be upper and lower bounded by [30]:

$$\overline{P}_n = \begin{cases} (1-\lambda)^{(1-\delta)M_A} & \text{if } n = 0 \\ \binom{(1+\delta)M_A}{n} \lambda^n (1-\lambda)^{(1+\delta)M_A - n} & \text{if } 1 \leq n \leq (1+\delta)M_A \\ 0 & \text{if } n > (1+\delta)M_A \end{cases} \quad (1a)$$

$$\underline{P}_n = \begin{cases} (1-\lambda)^{(1+\delta)M_A} & \text{if } n = 0 \\ \binom{(1-\delta)M_A}{n} \lambda^n (1-\lambda)^{(1-\delta)M_A - n} & \text{if } 1 \leq n \leq (1-\delta)M_A \\ 0 & \text{if } n > (1-\delta)M_A \end{cases} \quad (1b)$$

The value of $M_A$ depends on the specific source used and the parameters $\lambda$ and $\delta$ being optimized to provide the biggest secret key rate. Even if the source is UUS, the previous equations provide some bounds characterizing the photon number probabilities.

By fulfilling the condition $\lambda' q_A' = \lambda q_A$, both arrangements (passive and active) provide the same photon number distribution, because they have both the same source, the same internal losses and the same input power to the monitoring detector. Therefore, the upper and lower bounds of photon number probabilities for the passive arrangement can be estimated from those of the equivalent active set-up by means of Eqs. (1a, b) using $\lambda'$ instead of $\lambda$.

In the following, four cases will be analysed, depending on using infinite/finite keys and without/with decoy states.

## 2.2 Infinite key length

In the calculation of the secret key rate $R = \ell / N_A$ (secure bits, $\ell$, per pulse sent by Alice, $N_A$) is greatly simplified when the key string is considered as having an infinite length. In the context of UUS, the security analysis is carried out via untagged pulses [30]; nevertheless, Alice cannot measure (with current technology) their gain ($Q_u$) and the QBER ($E_u$), but the overall $Q$ and $E$ values. However, it is possible to provide some bounds based on the random sampling theorem. If the estimated upper bound of the probability for tagged pulses, acquired monitoring the pulses in the beam splitter ($q_A/1-q_A$), is $\overline{P}_t(\infty)$ ($\infty$ concerns the infinite key length), the lower bound for untagged pulses probability is $\underline{P}_u(\infty) = 1 - \overline{P}_t(\infty)$, and the upper and lower bounds for the gain and the gain times the error, are:

$$\overline{X}_u = X / \underline{P}_u(\infty) \text{ and } \underline{X}_u = \max\{0, (X - \overline{P}_t(\infty))/\underline{P}_u(\infty)\} \quad (2)$$

with $X \equiv Q, EQ$. Notice there is no distinction between $\underline{P}_u$ and simply $P_u$ when it is calculated for an infinite key length, so $\underline{P}_u(\infty) \equiv P_u(\infty)$, but there will be in the finite key case. The secret key rate for an infinite key length without using decoy states is:

$$R^\infty = q\{-Q f(E) h_2(E) + \underline{Q}_{1u}(1 - h_2(\overline{E}_{1u}))\} \quad (3)$$

$\underline{Q}_{1u} = \underline{Q}_u + \underline{P}_0 + \overline{P}_1 - 1$ being the lower bound for one-photon-untagged pulses and $\overline{E}_{1u} = Q\,E/\underline{Q}_{1u}$ its upper bound QBER; $\underline{P}_0$ and $\overline{P}_1$ are calculated using the Eq. (1a) and (1b). The $q$ value is the probability that Alice and Bob use the same basis in the measurement step, so in the infinite key length $q = ½$, $f(E)$ is the error correction inefficiency and $h_2$ is the usual binary Shannon information function.

The decoy method improves the performance of the QKD set-ups in the key-generation-rate as well as in the maximal secure distance $L_{max}$ (defined as the maximum transmission distance yielding a positive secure-key-rate). Alice assigns each bit randomly (or according to some probability $P_S$) to a signal or decoy state and both are attenuated internally with different transmittances $\lambda_{S,D}$. For signal and decoy pulses, normally $\lambda_D < \lambda_S$. The fundamental assumptions [33] of the decoy method are $Y_n^S = Y_n^D$ (the yield of an $n$-photon state for both the signal and decoy are equal) and $E_n^S = E_n^D$ (the QBER for both the signal and decoy states are equal). In this scenario, Eve only knows the output photon number ($n$) of each pulse. Unfortunately, if Eve has some knowledge about the source, the aforementioned main assumptions of the decoy state method fail [34]. This is the case of UUS, in which Eve knows the *output* ($n$) as well as the *input* photon number ($m$), being able to carry out an attack depending on both values. In this case the parameter that is the same is $Y_{m,n}$ (the conditional probability that Bob's detectors click given that the pulse enters Alice's set-up with photon number $m$ and is emitted with photon number $n$). Now the yield $Y_n = \Sigma_m P(m|n)\,Y_{m,n}$ ($P(m|n)$ being the conditional probability that a $m$-photon pulse enters Alice's set-up and is emitted as an $n$-photon pulse). As $P(m|n)$ depends on Alice's internal transmittances, being $\lambda_S \neq \lambda_D$, the yields will be different for signal and decoy states ($Y_n^S \neq Y_n^D$). In a similar way $E_n^S \neq E_n^D$, being $E_{m,n}^{S,D}$ (the QBER of pulses with $m$ input and $n$ output photons) for signal and decoy states, respectively.

For the case of infinite key length and using the weak + vacuum two state decoy protocol, the secret key generation rate is [30]:

$$R_D^\infty = q\,P_S \left\{ -Q^S f(E^S) h_2(E^S) + \underline{P}_u(\infty)\,\underline{Q}_{1u}^S (1 - h_2(\overline{E_{1u}^S})) \right\} \qquad (4)$$

Where $Q^S$ and $E^S$ are the overall gain and QBER of the signal states that can be measured experimentally; $P_S$ is the probability for a signal pulse to be sent (in the case of decoy random choosing, $P_S=1/2$) and $\underline{Q}_{1u}^S$ and $\overline{E_{1u}^S}$ are the lower and upper bound of the gain and the QBER, respectively, of the single photon states in untagged signal pulses, that can be estimated through the equations [30]:

$$Q_{1u}^S > \underline{Q}_{1u}^S = \frac{P_1^S\left( \underline{Q}_u^D \underline{P}_2^S - \overline{Q_u^S} \overline{P}_2^D + \overline{Q_u^V}(\overline{P_0^S} \overline{P}_2^D - \overline{P_0^D} \underline{P_0^S}) - \dfrac{2\delta M_A (1-\lambda_D)^{2\delta M_A - 1} \underline{P_2^S}}{((1-\delta)M_A + 1)!} \right)}{\overline{P_1^D}\,\underline{P_2^S} - \underline{P_1^S}\,\overline{P_2^D}} \qquad (5a)$$

$$E_{1u}^S \leq \overline{E_{1u}^S} = \frac{\overline{E_u^S}\,\overline{Q_u^S} - \underline{P_0^S}\,E_u^V\,Q_u^V}{\underline{Q}_{1u}^S} \qquad (5b)$$

$\underline{E_u^V} \underline{Q_u^V}$ being the lower bound of the error times the gain for the vacuum untagged pulses and can be evaluated by means of Eq. (2) as $\underline{E_u^V} \underline{Q_u^V} = \max(0, (E^V Q^V - \overline{P_t}(\infty)) / \underline{P_u}(\infty))$, as well as the remaining upper and lower bounded magnitudes related to the gain. In the infinite key length, $E^V = 1/2$. To calculate the probabilities, Eq. (1a) and (1b) should be used.

## *2.3 Finite key length*

The effect of using finite keys in the security bounds is not negligible [35]. The previous equations are for the secure key rate assuming their infinite length. However, the experimental set-ups are implemented for a finite period of time, providing finite key lengths. Strictly speaking, the length of the secret key generated in this case depends on the level of security required through externally imposed parameters. In this context, the concept of $\varepsilon$-security has been proposed.

A generic quantum key distribution protocol (without decoy states) has the following steps: the emitter (Alice) generates a number of $N_A$ pulses, from which the receiver (Bob) detects $N_B = Q N_A \leq N_A$ pulses ($Q$ being the gain). After the sifting ($N_S$) and the parameter estimation ($N_{PE}$) steps, the $N_B$ initially-shared qubit string is reduced to the raw key having $n = N_B - N_S - N_{PE} < N_B$ qubits, and finally, a classic post-processing (error correction and privacy amplification) produces the ultimate (smaller) secret key with length $\ell < n$. The secret key rate (defined here as secure bits per pulse sent by Alice) is $R = \ell / N_A = (\ell/n) (n/N_B) (N_B/N_A) = r q Q$ ($r$ being the secret fraction and $q \approx \frac{1}{2}$ for BB84).

To take into account the security under collective attacks within the framework of quantum adversaries, in which two classic systems (cc) are correlated with a quantum (q) one, the so-called "ccq-states formalism" [36] is needed. Representing the key space of $n$ bits as $\mathcal{K}$ and $K_A \equiv (k_A)^n$ and $K_B \equiv (k_B)^n$ (with $k_{A,B} \in \{0, 1\}$) being the individual keys held by Alice and Bob, Eve has the state $\rho_E^{K_A K_B}$ which is correlated with the keys $K_A, K_B \in \mathcal{K}$. The overall system can be described by the ccq-state

$$\rho_{K_A K_B E} = \sum_{K_A, K_B \in K} P(K_A, K_B) |K_A\rangle\langle K_A| \otimes |K_B\rangle\langle K_B| \otimes \rho_E^{K_A K_B},$$

$P(K_A, K_B)$ being a probability distribution defined in $\mathcal{K} \times \mathcal{K}$ and $K_A \neq K_B$ (in general). In this context, the definition for $\varepsilon$-security key has been introduced in [41, 15]:

Let $\rho_{K_A K_B E}$ be a ccq-state describing a classic pair of different keys ($K_A$, $K_B$) together an adversary holding the quantum system $E$. The pair of keys ($K_A$, $K_B$) is said to be $\varepsilon$-secure with respect to $E$, if and only if the condition $\frac{1}{2}\| \rho_{K_A K_B E} - \rho_{UU} \otimes \rho_E \|_1 \leq \varepsilon$ is fulfilled, where $\|.\|_1$ is the trace norm (or L$_1$-distance defined by means of this norm) and $\rho_{UU} \otimes \rho_E = \frac{1}{|\mathcal{K}|} \sum_{K \in \mathcal{K}} |K\rangle\langle K| \otimes |K\rangle\langle K| \otimes \rho_E$ is the ideal ccq-state, reflecting a uniformly distributed key (represented by the completely mixed state $\rho_{UU}$) and *uncorrelated* from the eavesdropper knowledge (represented by a tensor product of $\rho_E = tr_{K_A K_B}(\rho_{K_A K_B E})$ state).

The $\varepsilon$ parameter can be interpreted as the maximum probability that no secure key is generated [37, 38]. The security definitions based on a negligible accessible information do not imply universal security [39], especially in the context of quantum adversaries. However, the previous $\varepsilon$- secure security definition is universally secure [40,41] because the $L_1$-distance cannot increase under the action of any quantum operation [41]. If a key is $\varepsilon$-secure with respect to $E$, the mutual information between the key and $E$ is small, whereas the inverse is not, in general, true.

Using the previous definition of an $\varepsilon$-secure key, a new secret key rate can be established [38, 41] for finite key lengths. The starting point is a situation in which a correct key is shared $K_A = K_B = K$ and privacy amplification is carried out to get the final key of length $\ell$ fulfilling:

$$\ell \leq H_{\min}^{\bar{\varepsilon}}(\rho_{KE} | E) + 2\log_2(2/\varepsilon_{PA})$$

$\varepsilon_{PA}$ being the probability that privacy amplification fails, and $H_{\min}^{\bar{\varepsilon}}(\rho_{KE} | E)$ the quantum conditional $\bar{\varepsilon}$-smooth-min-entropy of $\rho_{KE}$ given $\mathcal{H}_E$ (eavesdropper's Hilbert space). The whole of Eve's information ($E$) can be separated into two pieces, $E \equiv "\mathcal{E} + \mathcal{C}"$, representing the information Eve can get by attacking the channel directly ($\mathcal{E}$) and through the additional data ($\mathcal{C}$) exchanged throughout the channel in the error correction process. In order to take into account the error probabilities coming from wrong error correction and parameter estimation, we have to proceed backward in the whole QKD protocol, from the starting point considered above [21].

The $\bar{\varepsilon}$-smooth-min-entropy can be bounded as $H_{\min}^{\bar{\varepsilon}}(\rho_{KE} | E) \equiv H_{\min}^{\bar{\varepsilon}}(\rho_{K\mathcal{E}\mathcal{C}} | \mathcal{E}\mathcal{C}) \geq H_{\min}^{\bar{\varepsilon}}(\rho_{K\mathcal{E}} | \mathcal{E}) - leak_{EC}$, the last term being the leakage of information in the error correction step $leak_{EC} \approx f(E) h_2(E) + \log_2(2/\varepsilon_{EC})$, and $\varepsilon_{EC}$ being the probability that the error correction step fails. The $\bar{\varepsilon}$-smooth-min-entropy term $H_{\min}^{\bar{\varepsilon}}(\rho_{K\mathcal{E}} | \mathcal{E})$ (defined as usual [42]) has to calculated, but the task is (in general) impossible. However, in the case of collective attacks, the state has a tensor product structure $\rho_{KE} = \rho_{(k\in)^n} = \sigma_{k\in}^{\otimes n}$, ($\in$ refers to the information extracted per qubit directly from the channel) allowing its bounding [41] by means of $H_{\min}^{\bar{\varepsilon}}(\rho_{K\mathcal{E}} | \mathcal{E}) \equiv H_{\min}^{\bar{\varepsilon}}(\sigma_{k\in}^{\otimes n} | \sigma_{\in}^{\otimes n}) \geq n(S(\sigma_{k\in} | \sigma_{\in}) - \delta(\bar{\varepsilon}))$ with $\delta(\bar{\varepsilon}) = 7((1 - \log_2 \bar{\varepsilon})/n)^{1/2}$, $\bar{\varepsilon} > 0$ being a parameter to be optimized and $\sigma_{\in} = tr_k(\sigma_{k\in})$. The conditional von Neumann entropy $S(\sigma_{k\in} | \sigma_{\in})$ has to be evaluated for a purification $\sigma_{k_A \approx k_B \in}$ of the approximate state $\sigma_{k_A \approx k_B} = tr_\in(\sigma_{k_A \approx k_B \in})$ such that were in the permitted set:

$$\Gamma_{\varepsilon_{PE}} = \left\{ \sigma_{k_A \approx k_B}, \left\| \lambda_m(\sigma_{k_A \approx k_B}) - \lambda_\infty(\sigma_{k_A \approx k_B}) \right\|_1 \leq \xi(\varepsilon_{PE}, m) \right\}.$$

For qubits, $\xi(\varepsilon_{PE}, m) = ((\ln(1/\varepsilon_{PE}) + 2\ln(m+1))/2m)^{1/2}$, $\lambda_m$ and $\lambda_\infty$ being the probability distributions obtained with $m$ (finite number) measurements and ideal (infinite number), respectively, on the states $\sigma_{k_A \approx k_B}$ compatible with the outcomes of the parameter estimation step and $\varepsilon_{PE}$ being its failure probability. Then $H_{\min}^{\bar{\varepsilon}}(\rho_{K\mathcal{E}} | \mathcal{E}) \leq n\left( \min_{\sigma_{k\in} \in \Gamma_{\varepsilon_{PE}}} S(\sigma_{k\in} | \sigma_\in) - \delta(\bar{\varepsilon}) \right)$.

Taking into account that the final secret key can only be generated by the one-photon untagged pulses, the minimum of the conditional von Neumann entropy can be rewritten [1] as $\underline{Q}_{1u}(1 - h_2(\overline{E}_{1u}))/Q$, $\overline{E}_{1u}$ being the upper bound for the QBER of the one-photon untagged pulses and $\underline{Q}_{1u}$ the lower bound for its gain.

Putting all the previous facts together, the achievable secure finite key rate is [38]:

$$R^F = q\{Q(-f(E)h_2(E)-\Delta) + \underline{Q_{1u}}(1-h_2(\overline{E_{1u}}))\} \tag{6}$$

with

$$\Delta = \frac{1}{n}\log_2\left(\frac{2}{\varepsilon_{PE}}\right) + 7\sqrt{\frac{1-\log_2\overline{\varepsilon}}{n}} + \frac{2}{n}\log_2\left(\frac{1}{2\varepsilon_{PA}}\right) \tag{7}$$

being the correction coming from the finiteness of the key.

Bearing in mind all of the process, the whole security parameter $\varepsilon$ for the final key has been split into several contributions, $\varepsilon = \varepsilon_{PA} + \varepsilon_{EC} + \overline{\varepsilon} + \varepsilon_{PE}$, some of these should be optimized (as will be shown in section 3) to provide the smallest $\varepsilon$ value and the largest secret key rate.

In the asymptotic limit of large (infinite) number pulses, Eq. (6) and (7) are in accordance with the Devetak-Winter bound [36] for the secret key rate, $R = S(K_A|Z) - H(K_A|K_B)$, $S$ and $H$ being the conditional von Neumann and Shannon entropies, respectively, evaluated for the joint state of Alice and Eve and for Alice and Bob.

The Eq. (6) and (7) can be used in the context of decoy states together with the same $\Delta$ correction and the estimation of $\underline{Q_{1u}}$ and $\overline{E_{1u}}$ by means of the Eq. (5a) and (5b), including the suitable finite data fluctuation in the estimation of $\underline{P_u}$. The secure finite key rate is:

$$R_D^F = q\,P_S\{Q^S(-f(E^S)h_2(E^S)-\Delta) + \underline{P_u}\,\underline{Q_{1u}^S}(1-h_2(\overline{E_{1u}^S}))\} \tag{8}$$

**3 Numerical simulation of secure key rates**

In the following, a numerical simulation using the set-up shown in figure 1, will be carried out. Four situations will be considered: without/with decoy states together infinite/finite key length.

The "plug-and-play" set-up assumes a UUS situation. Therefore, Alice should not suppose any photon distribution from Bob's source but, in order to be able to calculate the main characteristics of the pulses, an outgoing Poissonian distribution is assumed to be released by Bob, with average photon number $M_B$. In all cases addressed, a passive set-up has been considered. Consequently, after the channel attenuation (characterized by the loss coefficient $\beta$), Alice's beam splitter $q_A/1-q_A$ and internal attenuation (characterized by transmittance $\lambda$ when no decoy states are used or by $\lambda_{S,D}$ if signal and decoy states are included), the average photon number for the pulses sent to Bob is $\mu = M_A \lambda q_A = M_A \lambda' q_A' = M_A \lambda' (1-q_A)$ (or $\mu_{S,D} = M_A \lambda_{S,D} q_A$ with decoy states) with $M_A = M_B\,10^{-\beta L/10}$. In plug-and-play systems, the distance $L$ is considered as the spatial separation between Alice and Bob, even though a round trip of the pulses is taken into account in order to calculate the real losses. Depending on the experimental strategy used, different lower and upper bounds for the gain and QBER will be considered depending on the $\underline{P_u}$ estimation.

Some values for the parameters used in the following simulation are taken from [43] ($\eta_B = 0.045$, $\beta = 0.21$ dB/km, $Y_0 = 1.7\,10^{-6}$, $e_{det} = 0.033$) and the remaining ones are chosen as representative values ($M_B = 10^6$, $q_A = 0.01$, $f(E) = 1.22$, $E_0 = 0.5$, $E_0^V = 0.5$, $\varepsilon_{EC} = 10^{-10}$ and the total security parameter of the generated keys $\varepsilon = 10^{-9}$).

*3.1 Without decoy states*

In this paragraph the effect of a finite key length in the secure key rate is considered, when decoy states are not used. Similar results have already been reported in [30] with infinite key length and are recalculated here in order to compare the effect of the finite key case. The simulation uses Eq. (3) and (6). Assuming a Poissonian photon distribution for Bob's source (with average photon number $M_B$), Alice's source is also Poissonian with average photon number $\mu = M_A \lambda q_A$. The overall gain ($Q$) and QBER ($E$) can be calculated by means of $Q = Y_0 + 1 - e^{-\mu\eta}$ and $EQ = E_0 Y_0 + e_{det}(1-e^{-\mu\eta})$, and the untagged pulse probability is $\underline{P_u}(\infty) = erf(\delta\sqrt{M_A(1-q_A)/2})$, $Y_0$ and $E_0$ ($\approx 0.5$) being the background yield and its error, and $e_{det}$ the intrinsic error rate for the detector. The parameters $\lambda$ and $\delta$ are numerically optimized to provide the largest key rate. As is discussed in [31] the efficiency of a passive strategy is improved if more photons are sent to the intensity monitor, meaning a small $q_A$ value. Here $q_A = 0.01$ and $M_B = 10^6$ are taken as reasonable values and the results of $R^\infty$ are in accordance with those published in [31] (see figure 2a).

Now the finite-key case is addressed. As was mentioned before, Alice sends Bob a number of $N_A$ pulses, from which the receiver (Bob) detects $N_B \leq N_A$. In fact, these $N_A$ pulses come from Bob's source (in a previous round trip) with the initial average photon number $M_B$, attenuated by the channel losses, and are sent back to Bob. The raw key has length $n = N_B - N_S - N_{PE} < N_B$, $N_S$ and $N_{PE}$ being the number of pulses used in the sifting and parameter estimation processes, respectively. To conclude, a classic post-processing (error correction and privacy amplification) produces the final secret key length $\ell < n$. The secret key rate is calculated using Eq. (6) and (7). However, now to estimate $\underline{Q_u}$, a finite number ($N_A$) of untagged pulses has to be taken into account, so the distribution probability of untagged pulses ($P_u(N_A)$) will have a deviation characterized by the condition $|P_u(N_A) - P_u(\infty)| \leq \xi(\varepsilon_u, N_A)$. The new lower and upper bounds for the untagged and tagged pulses probabilities are $\underline{P_u}(N_A) = P_u(\infty) - \xi(\varepsilon_u, N_A)$ and $\overline{P_t}(N_A) = 1 - \underline{P_u}(N_A)$, and using Eq. (2):

$$\underline{Q_u} = \max\left(0, \frac{Q - P_u(\infty) - \xi(\varepsilon_u, N_A)}{1 - P_u(\infty) - \xi(\varepsilon_u, N_A)}\right) \qquad \underline{E_{1u}} = \frac{Q(E + \xi(\varepsilon_E, m_E))}{\underline{Q_{1u}}} \qquad (9)$$

The total security parameter for the finite key QKD protocol is $\varepsilon = \varepsilon_{PA} + \varepsilon_{EC} + \bar{\varepsilon} + \varepsilon_u + \varepsilon_E$, showing several error contributions: $\varepsilon_{PA}$ in privacy amplification, $\varepsilon_{EC}$ is the error in the error correction, $\bar{\varepsilon}$ bounding the $\bar{\varepsilon}$-smooth-min-entropy term and two contributions to the parameter estimation term, one coming from the estimation of untagged-pulses probability ($\varepsilon_u$) and another from the QBER ($\varepsilon_E$) estimation, with the deviations of $\xi(\varepsilon_u, N_A)$ and $\xi(\varepsilon_E, m_E)$, respectively; $m_E$ being the number of pulses used to estimate the QBER. The finite secret key rate $R^F$ is calculated considering fixed typical values for $\varepsilon$, $\varepsilon_{EC}$, $q_A$ and $M_B$ and optimizing the remaining parameters {$\varepsilon_{PA}$, $\bar{\varepsilon}$, $\varepsilon_u$, $\varepsilon_E$, $\lambda$, $\delta$, $m_E$} to get a maximum secret key rate. Note that $N_{PE}$ is the number of pulses used in the parameter estimation processes, and the present passive set-up does not decrease the number of pulses in the untagged-probability estimation process [31]. As a consequence, the only step that removes pulses in the parameter estimation comes from the QBER, then $N_{PE} \equiv m_E$.

In figure 2a, the effect of the finiteness on the secret key rate is shown. Only when $N_A \geq 10^{14}$ the $R^F \approx R^\infty$. Assuming $N_B - N_S \approx N_B/2$ (then $n = N_B - N_S - N_{PE} = N_B/2 - m_E$) the number of pulses $m_E$ used to estimate the QBER should increase with the distance $L$. Figure 2b show the relationship $r =$

$m_E/(N_B/2)$ as representing the optimum percentage of sampling qubits with respect the sifted key ($\sim N_B/2$) required to estimate the QBER. For instance, using a set-up with the previously specified parameters, for $N_A = 5 \; 10^{10}$, the sifted key has $N_B/2 \approx 4 \; 10^6$ bits, $m_E \approx 7 \; 10^5$, then an $r_S \equiv 18\%$ of the sifted key should be sacrificed in order to reach the maximum secure distance of $L_{max} \approx 20$ km having $R^F \approx 2 \; 10^{-6}$. Obviously, by keeping the same $R^F$, increasing the number of pulses until $N_A = 10^{14}$, it would possible to reach $L_{max} = 33$ km, sacrificing only 1.5% of the sifted key. Figure 2c shows the optimal average photon number per pulse $\mu = M_A \, \lambda \, q_A$ as a function of $L$. Only the results for $N_A = 5 \; 10^{10}$ are shown because for a bigger $N_A$ the curves are closer to that of the infinite case.

In order to see the L-threshold for a positive $R^F$ value, the maximum secure distance ($L_{max}$, in km) fulfilling the (reasonable mathematical) condition $R^F \approx 10^{-9}$ is shown in figure 3 (lower curve), versus the number of pulses sent by Alice ($\log N_A$). In order to provide a threshold for positive values of the secure key rate, the equation $L_{max}(R^F \approx 10^{-9}, N_A) = 0$, is solved to obtain a value of $N_A = N_A^{th} \approx 10^9$. Therefore, for $N_A \geq N_A^{th}$, a positive value of the secret key rate will be obtained and the whole QKD process is secure, assuming the values of the remaining parameters constant.

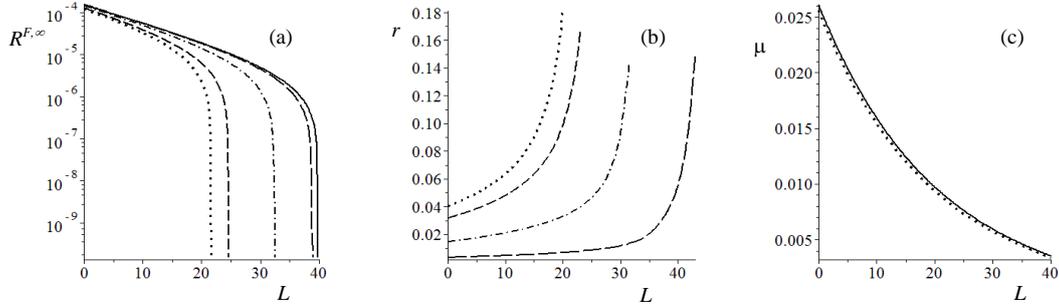

Fig. 2. (a) Secure key rate versus $L$ when no decoy states are used ($R^{F,\infty}(L)$): (solid line) infinite key length, non-solid lines are for several key lengths: $N_A = 5 \; 10^{10}$ (dotted line), $N_A = 10^{11}$ (dashed line), $N_A = 10^{12}$ (dot-dash line), and $N_A = 10^{14}$ (long dashed line). (b) Relationship $r = m_E/(N_B/2)$ versus $L$ for $N_A = 5 \; 10^{10}, 10^{11}, 10^{12}, 10^{14}$ (the same lines as before are used). (c) Optimal average photon number per pulse ($\mu$) for: infinite key (solid line) and $N_A = 5 \; 10^{10}$ (dotted line).

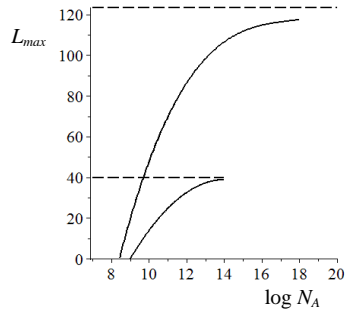

Fig. 3. Maximum secure distance ($L_{max}$ in km) as a function of $\log N_A$ for: upper curve, with the condition $R_D^F = 10^{-9}$, lower curve $R^F = 10^{-9}$. The asymptotic straight (dashed) lines are the limits for $N_A = \infty$ calculated with the condition $R^\infty$ or $R_D^\infty = 10^{-9}$: upper line providing $L_{max} \approx 123$ km and lower line providing $L_{max} \approx 40$ km.

## 3.2 With vacuum+weak decoy states

In this paragraph the effect of a finite key length in the secure key rate is considered, when decoy states are used. Some results have already been reported by [30] for infinite key length and are recalculated here in order to compare the effect of the key finiteness.

The infinite key length ($\Delta = 0$) simulation uses Eq. (4). The overall gain and QBER are given by $Q^{S,D} = Y_0 + 1 - e^{-\mu_{S,D}\eta}$ and $E^{S,D} Q^{S,D} = E_0 Y_0 + e_{det}(1 - e^{-\mu_{S,D}\eta})$, $\mu_{S,D} = M_A \lambda_{S,D} q_A$ being the signal and decoy pulses intensities. The protocol assumes a strategy in which the signal and decoy states are chosen randomly, then $P_S = ½$ in Eq. (4). The values of $\underline{Q_{1u}^S}$ and $\overline{E_{1u}^S}$ are calculated by means of Eq. (5a) and (5b) and $\overline{P_t(\infty)} \equiv \underline{P_t(\infty)} = erfc(\delta\sqrt{M_A(1-q_A)/2})$. The parameters $\lambda_{S,D}$ and $\delta$ are numerically optimized to maximize the key rate (Eq. (4)). As previously assumed, $q_A = 0.01$ and $M_B = 10^6$ are taken as reasonable values. The results of $R_D^\infty$ are in accordance with those published in [30].

As the random signal-pulse strategy (with $P_S = ½$) is not the optimal method, the probability $P_S$ should be optimized fulfilling the condition $P_S + P_D + P_V = 1$, $P_{S,D,V}$ being the probabilities for the signal, decoy and vacuum pulses to be chosen by Alice. Therefore, for each value of $L$, Alice has $N_A$ initial pulses (received from Bob), from which she chose $N_A^S = N_A P_S$ and $N_A^D = N_A P_D$ as signal and decoy states. After Bob's detections, they share $N_A P_S Q^S$ and $N_A P_D Q^D$ detected pulses. Once the sifting and parameter estimation have been carried out, the raw key signal states has length $n = N_A P_S Q^S / 2 - m_E^S$, $m_E^S$ being the signal pulses used to estimate the error and that should be optimized for each $L$.

To calculate the secure key rate effect, Eq. (8) and (7) are used. The bounded magnitudes for signal and decoy states can be calculated by means of Eq. (2). Now the appropriate finite fluctuation has to be included in several places. In order to calculate the values for $\underline{Q_u^D}, \overline{Q_u^S}$ and $\overline{Q_u^V}$, the fluctuations are incorporated by means of the untagged probabilities $\underline{P_u^i} = P_u(\infty) - \xi(\varepsilon_u^i, N_A^i)$, $i = S, D, V$, affected by the errors $\varepsilon_u^i$. In addition, the signal error estimation introduces a new error as $\overline{E^S} = E^S + \xi(\varepsilon_E^S, m_E^S)$, $m_E^S$ being the number of pulses used to estimate the QBER of signal pulses.

Again, the total security parameter for this finite key QKD protocol is $\varepsilon = \varepsilon_{PA} + \varepsilon_{EC} + \bar{\varepsilon} + \varepsilon_u^S + \varepsilon_u^D + \varepsilon_u^V + \varepsilon_E^S$, showing three new error contributions to the parameter estimation coming from the probability of untagged-pulses and another from the QBER ($\varepsilon_E^S$). The final secret key rate $R^F$ is calculated considering fixed typical values for $\varepsilon$, $\varepsilon_{EC}$, $q_A$ and $M_B$ and optimizing the remaining parameters {$\varepsilon_{PA}, \bar{\varepsilon}, \varepsilon_u^S, \varepsilon_u^D, \varepsilon_u^V, \varepsilon_E^S, \lambda_S, \lambda_D, \delta, m_E^S, P_S, P_D, P_V$} to get a maximum secret key rate. In spite of all the parameters being optimized, as already mentioned in [24], the secure key rate values are almost independent of {$\varepsilon_{PA}, \bar{\varepsilon}, \varepsilon_u^S, \varepsilon_u^D, \varepsilon_u^V$} when they are changed within a rational range. The most important parameters are the total number of pulses used and the signal-to-decoy states relationship.

Figure 5a displays the results for $R_D^\infty$ and $R_D^F$, showing a significant increase in the maximal secure distance $L_{max}$ with respect to those obtained without using decoy states. This is also reflected in figure 3 (upper curve). With this strategy, and the parameters used, a number of $N_A \geq 10^{16}$ pulses should be used in order to reach a secure key rate near the $R_D^\infty$. It is remarkable that in the short and medium distances region, a higher key rate is obtained by means of a finite number of pulses if the probabilities $P_{S,D,V}$ are optimized instead of taking $P_S = ½$. In order to see the L-threshold for a positive $R_D^F$ value, the maximum secure distance ($L_{max}$, in km) fulfilling the condition $R_D^F \approx 10^{-9}$, is shown in figure 3 (upper curve), versus the number of pulses sent by Alice ($\log N_A$). In order to provide a threshold for the positive values of the secure key rate, the equation $L_{max}(R_D^F \approx 10^{-9}, N_A) = 0$, is solved to obtain a value of $N_A = N_A^{th} \approx 3\ 10^8$. Therefore, for $N_A \geq N_A^{th}$, a positive value of the secret key rate will be obtained and the whole QKD process is secure assuming the values of the remaining parameters constant. Decoy states decrease the threshold needed to get positive secret key rates in almost one order of magnitude.

Figure 5b shows the optimized probabilities versus $L$ (distance between Alice and Bob). The value of $P_D$ increases with $L$, indicating that more resources are needed to estimate the error bounds. In the limit of quasi-infinite key length ($N_A \geq 10^{16}$), almost all the pulses should be considered as signals. The probabilities are near ½ only in the long $L$ region. Out of this $L$ range, sending more signals than decoy states provides higher secure key rate.

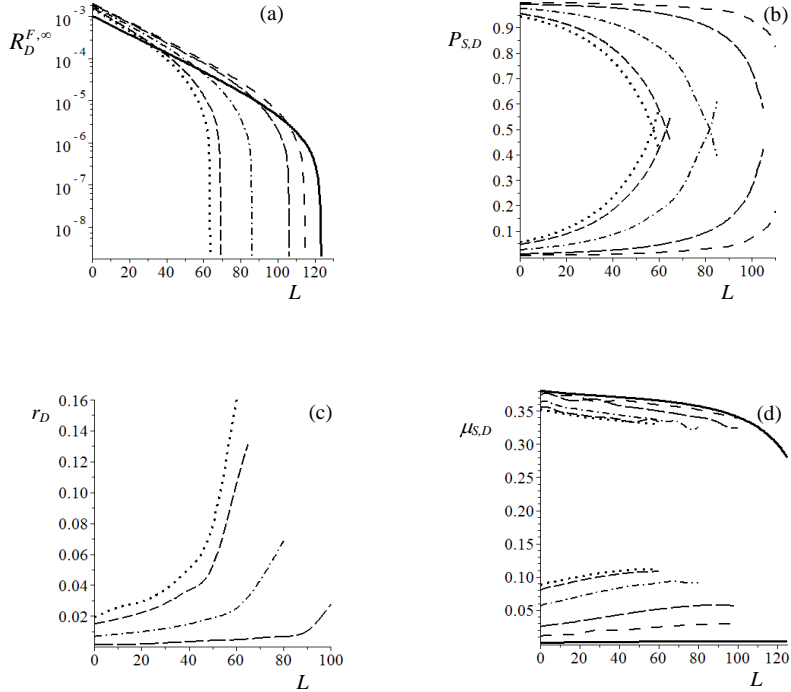

Fig. 5. (a) Secure key rate versus $L$ when decoy states are used ($R_D^{F,\infty}(L)$), for: $N_A = 5\ 10^{10}$ (dotted line), $10^{11}$ (dashed line), $10^{12}$ (dot-dash line), $10^{14}$ (long dashed line), $10^{16}$ (space dashed line) optimizing $P_S$, $P_D$ and $P_V$, and infinite key length (solid line). (b) Optimized probabilities $P_S$ (upper curves), $P_D$ (lower curves) versus $L$ (with the same line notation). (c) $r_D = m_E^S/(N_A P_S Q^S/2)$ relationship versus $L$ for $N_A = 5\ 10^{10}, 10^{11}, 10^{12}, 10^{14}$ (with the same line notation). (d) Signal and decoy intensity pulses versus $L$ (with the same line notation).

The raw key is $n = N_A P_S Q^S/2 - m_E^S$, and figure 5c shows the relationship $r_D = m_E^S/(N_A P_S Q^S/2)$ as representing the optimum percentage of sampling qubits with respect to the sifted signal key ($N_A P_S Q^S/2$) required to estimate the QBER. The number of pulses $m_E^S$ used to estimate the QBER increases with the distance $L$. In order to compare the effect of including decoy states, for the case of $N_A = 5 \ 10^{10}$ and $L_{max} \approx 20$ km (maximum secure distance without decoy states), now the sacrificed sifted key is $r_D \equiv 3\%$ and $R_D^F = 4 \ 10^{-4}$, a value two orders of magnitude larger than that obtained without decoy states. On the other hand, to have a value of $R_D^F \sim 4 \ 10^{-6}$ with $N_A = 5 \ 10^{10}$ using decoy states, the sacrificed sifted key is rate $r_D \sim 16\%$ but $L_{max} \approx 60$ km.

Figure 5d details the signal and decoy pulsed intensities versus $L$. The intensities of decoy pulses increase with $L$ reflecting that larger resources for key distribution are needed. In addition, their intensities are higher than those obtained in the infinite key length case.

Finally, in order to identify the terms that have the largest losses due to the finite key effect, tests have been carried out including decoy states. As an example, Eqs. (4) and (8), have been checked for both infinite and finite key length, respectively. For the latter case, $N_A = 5 \ 10^{10}$ (see figure 5a) is considered, the relationship being $R_D^\infty / R_D^F$ ($L=60$) ~ 10. The first term in (8) (related to the error correction process) includes a $\Delta$ correction that is negligible, because the errors involved ($\varepsilon_{PE}, \bar{\varepsilon}, \varepsilon_{PA}$) are affected by a logarithm and, always, divided by n (the raw key length). As a consequence, this term does not give rise to any significant differences in the key rates. The second term (related with the privacy amplification process) in both equations has a crucial behaviour as its value in $R_D^F$ is around 25 times smaller than in $R_D^\infty$. The origin of this difference is located in the upper bound of the quantum-bit-error-rate of the signal pulses, because (for $L = 60$ km) $\overline{E_{1u}^S}(finite\,key,\,Eq.(8)) \approx 5 \ \overline{E_{1u}^S}(infinite\,key,\,Eq.(4))$, even increasing dramatically for $L > 60$. This is also reflected in the values of $\underline{Q_{1u}^S}$, where $\underline{Q_{1u}^S}(finite\,key,\,Eq.(8)) \approx \underline{Q_{1u}^S}(infinite\,key,\,Eq.(4))/5$.

## 4 Conclusions

The security of the plug-and-play passive set-up has been analysed for the BB84 QKD protocol when the effect of finite key lengths is included. An unknown and untrusted source scenario is assumed, reflecting a situation in which the eavesdropper could tamper with the source and the channel completely. For the values of the parameters used, this arrangement is $\varepsilon$-secure for a wide range of pulse number. Using decoy states enlarges the maximal secure length significantly as well as the secure key rates for all the number of pulses considered. It is particularly noticeable that using finite key lengths can provide larger key rates if the rate between the number of signal to decoy states is optimized. This behaviour appears in the short and medium distance region.

In spite of the numerical simulation involving the optimization of the main parameters, the results conclude that the most important parameters are the total number of pulses used and the signal-to-decoy-state relationship. The parameters related to the error correction process (the first term in the key rate) has little effect on the final key rates. The main influence resulting from the finite key length concerns the upper bound of the quantum-bit-error-rate of signal pulses. This directly affects the term coming from the privacy amplification process and in the lower bound of one-photon untagged pulses.

**Acknowledgments**

The author would like to thank V. Martín and the Research Group on Quantum Information and Computation at the Technical University of Madrid (UPM) for the interesting discussions and suggestions about the subject. This work has been supported by the research project Quantum Information Technologies (QUITEMAD), P2009/ESP-1594 of Comunidad Autónoma de Madrid in Spain.